# On the rearrangement time of the fission reaction

G. Mouze and C. Ythier
Faculté des Sciences, Université de Nice, France

The rearrangement time Δt of the fission reaction can be extracted from the full-width at half maximum (f.w.h.m.) of the isotopic distributions of fission fragments if this width is attributed to an uncertainty ΔN in the neutron-number N of the fragment; then the energy-time uncertainty relation leads to Δt = 0.17 yoctosecond.

In the fifties, data obtained from a great number of spontaneously fissioning nuclei and from $^{233}$U, $^{235}$U and $^{239}$Pu irradiated with 80 keV neutrons were used for the determination of the probability P(ν) of emitting ν neutrons per fission. The representation of P(ν) as a function of the difference $\nu - \bar{\nu}$ , where $\bar{\nu}$ is the average number of neutrons emitted per fission, is a bell-shaped curve, and J. Terrell [1] demonstrated in 1957 that it can be fitted by a gaussian curve having a σ –parameter equal to 1.08, what corresponds to a f.w.h.m. of 2.538 u for P(ν) = f ($\nu - \bar{\nu}$ ).
Terrell tried to explain this situation as resulting from large variances in total kinetic energy and hence in excitation energy of the fragments.

In the nineties, the study, using coincidence methods and detectors such as Gammasphere, of the prompt gamma rays emitted by neutron –rich fission fragments revealed that the *isotopic* distributions of fission fragments can be represented by gaussian curves [2].

In 1996, J.L. Durell [3] pointed out that all isotopic distributions encountered in fission have the same f.w.h.m. as that found by J. Terrell for the P(ν) –distribution. In his study of the gamma-rays emitted in the fission of $^{248}$Cm, he obtained the isotopic distributions of the Zr- fragments associated with each of the Ba-fragments of the spontaneous fission of $^{248}$Cm. In his figure 7, the solid curves are precisely the gaussian distributions "expected" in fission. He proposed the same explanation as Terrell.

In our opinion, the fact that all isotopic distributions of fission fragments have a width of 2.54 u is the manifestation of the energy-time uncertainty relation, and of the extreme brevity of the main step of the fission reaction, which causes an uncertainty in mass given by: $\Delta A = \hbar/(c^2 \Delta t)$, where Δ t is extremely small. We proposed for the first time in 1991 [4] the idea that the extreme brevity of *the cataclysmic rearrangement* of nuclear fission could be put into evidence by an uncertainty ΔA in the mass A of the fission fragments; a ΔA- value of about 5 u, corresponding to a Δt –value of $1.410^{-25}$s was then justified. The discovery, by C. Signarbieux [5] of a fine structure in the mass distribution of fission fragments allowed in 1993 a more precise determination of ΔA: a fine structure was observed in the region A = 104 of the mass distribution of the neutron-induced fission of $^{235}U$ at the highest value of the total kinetic energy of the fragments. But the mass- line at A = 104 corresponds to the light fragment, $^{104}Mo$, of the most energy-rich fragment pair $^{104}Mo$- $^{132}Sn$ (Q = 205.88 MeV), because no other light fragment is expected in this mass region at this energy; consequently, the mass- lines at A = 105, 106 and 107 cannot belong to other light fragments, and all the satellite lines around A = 104 have to be considered as the effect of the transformation of the discrete mass-value A = 104 into a broad mass-line, centred on A = 104 : indeed, in 1993, our analysis of this mass spectrum led to a gaussian distribution having a width of 4 u [ 6 ].

Similar cold-fission data reported by F.Gönnenwein [7] lead, if they are appropriately analysed, to an even more precise value of ΔA, namely ΔA = 4.15 u, thanks to a better statistics.

It is noteworthy that the values 4.15 u and 2.538 u are in a ratio of A/N, since 4.15/2.538 = 1.6351, whereas the mean value of A/N between $^{236}U$ and $^{252}Cf$ is equal to 1.6376. This situation constitutes a strong indication that the width of the isotopic distributions and that of the mass distributions have the same origin, namely the extremely small value of the reaction time!

According to this hypothesis, the mass-width ΔA of fission fragments can now be deduced from the Δ N-value of Terrell's law:

$\Delta A = \overline{(A/N)}\, \Delta N = 1.6376 \times 2.538\ u = 4.156\ u$,

i.e. be based on the considerable work made in the fifties for determining $\bar{\nu}$ and P(ν) of the light actinide nuclei, data which are of the utmost practical importance in the application of the chain reaction in reactors or explosives !

As ΔA = ΔZ + ΔN, ΔZ, the uncertainty in the charge of the fragments, is given by:

ΔZ= ΔA - ΔN = 1.618 u;

and the most precise Δt – value is now:

$\overline{\Delta t} = \hbar/(c^2 \times 4.156 u) = 1.70\ 10^{-25}$ s.

Consequently, to any fission fragment is attached

--an uncertainty in A, ΔA = 4.156 u,

--an uncertainty in N , ΔN =2.538 u,

--an uncertainty in Z, ΔZ = 1.618 u.

These uncertainties are the "trademark" of the fragment, a sign that it was formed in the "nucleon phase" [8]. And we may conclude that the energy-time uncertainty relation explains, at the same time, the law of Terrell concerning the P(ν) - distribution, the law of Durell of the isotopic distributions encountered in fission, and the law of distribution of the charge in fission. In fact, neither $\bar{\nu}$, nor the N-value corresponding to the maximum yield in isotopic distributions are integer numbers.

It must be pointed out that a reaction time of 0.17 yoctosecond is not surprising for a reaction occurring within an atomic nucleus. If the reaction time of a chemical reaction can be defined as the ratio of range $l$ of interaction to velocity of propagation of the interaction, its smallest value is $l$/c, i.e, for a reaction occurring between H and Cl in a HCl molecule, ~ 0.43 $10^{-18}$ s, with $l$ ~ 1.27455 $10^{-10}$ m [9 ]. And for a reaction occurring within the most external valence shells of a nucleus, $l$ can be about $10^{7}$ times smaller, and Δt can be of the order of 1 $10^{-25}$ s. Interestingly, the value Δt = 0.17 ys corresponds to a range of ~5.09 $10^{-17}$m.

In conclusion, the energy-time uncertainty relation allows, as it is well-known from high-energy physics, an estimation of extremely short time intervals through the corresponding energy-width; but in fission we are confronted, more precisely, by mass-widths, neutron-widths and proton-widths. The value Δt = 0.17 yoctosecond, justified in the present paper, corresponds to *extreme conditions*, which are those of the "nucleon phase" in which *the sharing out* of "nucleons" between the nascent light and heavy fragments occurs [8,10,11].

------------------------------------